\documentclass[aps,twocolumn,showpacs,preprintnumbers,amsmath,amssymb,superscriptaddress]{revtex4-2}
\usepackage{graphicx}
\usepackage{dcolumn}
\usepackage{bm}
\usepackage{color}
\usepackage{amsmath}
\usepackage{amssymb}
\usepackage{latexsym}
\usepackage{multirow}
\usepackage{xcolor}
\usepackage{ulem}
\usepackage{hyperref}
\usepackage[capitalise]{cleveref}
\usepackage{hyperref}
\usepackage{mathtools}
\newcommand\underrel[3][]{\mathrel{\mathop{#3}\limits_{
			\ifx c#1\relax\mathclap{#2}\else#2\fi}}}
\begin{document}
	\title{Numerical simulations of confined Brownian-yet-non-Gaussian motion}
	\author{Elodie Millan}\thanks{These authors contributed equally.}
	\affiliation{Univ. Bordeaux, CNRS, LOMA, UMR 5798, F-33400, Talence, France.}
	\author{Maxime Lavaud}\thanks{These authors contributed equally.}
	\affiliation{Univ. Bordeaux, CNRS, LOMA, UMR 5798, F-33400, Talence, France.}
	\author{Yacine Amarouchene}\email{yacine.amarouchene@u-bordeaux.fr}
	\affiliation{Univ. Bordeaux, CNRS, LOMA, UMR 5798, F-33400, Talence, France.}
		\author{Thomas Salez}\email{thomas.salez@cnrs.fr}
	\affiliation{Univ. Bordeaux, CNRS, LOMA, UMR 5798, F-33400, Talence, France.}
	\begin{abstract}
Brownian motion is a central scientific paradigm. Recently, due to increasing efforts and interests towards miniaturization and small-scale physics or biology, the effects of confinement on such a motion have become a key topic of investigation. Essentially, when confined near a wall, a particle moves much slower than in the bulk due to friction at the boundaries. The mobility is therefore locally hindered and space-dependent, which in turn leads to the apparition of so-called multiplicative noises, and associated non-Gaussianities which remain difficult to resolve at all times. Here, we exploit simple, optimized and efficient numerical simulations to address Brownian motion in confinement in a broadrange and quantitative way. To do so, we integrate the overdamped Langevin equation governing the thermal dynamics of a negatively-buoyant single spherical colloid within a viscous fluid confined by two rigid walls, including surface charges. From the produced large set of long random trajectories, we perform a complete statistical analysis and extract all the key quantities, such as the probability distributions in displacements and their main moments. In particular, we propose a novel method to compute high-order cumulants by reducing convergence problems, and employ it to efficiently characterize the inherent non-Gaussianity of the confined process.
\end{abstract}
\maketitle

\section{Introduction} 
Brownian motion is the random movement of a microparticle due to thermal agitation. This name was given in reference to Robert Brown, a botanist, who observed in 1828 for the first time the erratic trajectories of pollen grains, using a microscope \cite{brownBriefAccountMicroscopical1828}. He concluded that this motion was not from a living source since he observed the same phenomenon with grains of inorganic matter, like minute fragments of window glass or mineral substances. In 1905, Albert Einstein \cite{einsteinUberMolekularkinetischenTheorie1905}, William Sutherland \cite{sutherlandDynamicalTheoryDiffusion1905} and Marian Von Smoluchowski \cite{vonsmoluchowskiZurKinetischenTheorie1906a}, physically modelled Brownian motion, and independently calculated the diffusion coefficient of a single particle, assuming that matter is discontinuous. In 1909, Jean Perrin \cite{perrinMouvementBrownienGrandeurs1909} validated Einstein's theory by studying both the distribution and the agitation of microparticles in suspension. Moreover, doing so, he experimentally measured the Avogadro number, proving the atomic nature of matter, and was thus awarded the Nobel Prize in 1926. Besides, from this work, it then became clear that Brownian motion can be a probe of conservative forces. In 1908, Paul Langevin \cite{langevinTheorieMouvementBrownien1908} developed the equation that governs Brownian trajectories using the fundamental principle of dynamics, and taking into account both the viscous Stokes force and a new stochastic force representing the effect of momentum transfer from collisions with solvent molecules. It is worth stressing that many variations and applications around Brownian motion in the bulk are continuously explored nowadays, and some of the key underlying hypotheses and concepts remain at the heart of epistemological discussions ~\cite{Genthon2020,Ardourel2022}. 

In the second half of the 20th century, the rise of miniaturization triggered the need for a further understanding of interfacial and confinement effects on colloidal mobility~\cite{brennerSlowMotionSphere1961,batchelorBrownianDiffusionParticles1976,Bian2016}. Almost ninety years after the Einstein-Sutherland-Smoluchowski theory, a seminal study of Brownian motion near rigid walls was performed~\cite{faucheuxConfinedBrownianMotion1994}. In the latter, the space-dependent wall-friction-induced reduction in the average planar diffusion coefficient
of confined colloids was revealed. This result triggered a novel research activity on Brownian motion in confinement \cite{Dufresne2000,Felderhof2005,Joly2006,Eral2010,Benichou2013,Mo2015,matseTestDiffusingdiffusivityMechanism2017,Gubbiotti2019}, with implications in single-molecule force spectroscopy~\cite{Strick1996,Bockelmann2002}, and surface-force measurements~\cite{Sainis2007,lavaudStochasticInferenceSurfaceInduced2021}. Random and active motion of microalgae near boundaries may also be impacted by the existence of an altered mobility in confinement~\cite{Rieu2021}. A key feature of confined Brownian motion is the emergence of multiplicative noises due to the space-dependent diffusion constants near walls. A direct implication of such noises is the non-Gaussianity of the particle's displacement distributions~\cite{han2006brownian,munk2009effective,wang2009anomalous,leptos2009dynamics,wang2012brownian,skaug2013intermittent,chubynsky2014diffusing,guan2014even,jain2016diffusion,jain2016diffusingsurvival,chechkinBrownianNonGaussianDiffusion2017,sposini2018first,lanoiselee2018diffusion,lanoiselee2018model,czajka2019effects,chakraborty2020disorder,hidalgo2020hitchhiker,yin2021non,miotto2021length,alexandreNonGaussianDiffusionSurfaces2022}, despite the mean-square displacements (MSDs) remain linear in time (as expected for a classical Brownian process). 
Moving beyond rigid confinement at equilibrium, the influence of fluctuating interfaces on Brownian point-like tracers was
investigated theoretically~\cite{Marbach2018}, and experimentally~\cite{Sarfati2021}, as well as the effects of wall adhesion~\cite{Benichou2008,Boniello2018,Alexandre2022}. Besides, fluid and soft boundaries were considered~\cite{Bickel2006,Wang2009,Daddi2016,Bar2017}, and Taylor dispersion in confinement was investigated \cite{Cejas2017,Vilquin2021,Vilquin2022}.

While simple Brownian motion has been numerically modelled in the bulk (see Ref.~\cite{volpeSimulationBrownianParticle2013} for a tutorial), as well as in more complicated cases involving confinement and interactions \cite{petersEfficientBrownianDynamics2002b,simonninDiffusionConfinementHydrodynamic2017,sprinkleDrivenDynamicsDense2020, knowlesCurrentFluctuationsNanopores2021, marbachIntrinsicFractionalNoise2021}, an efficient and quantitative numerical approach allowing for a broadrange characterization of non-Gaussianities, is lacking to date. In this article, we aim at filling this gap. We describe how one can model the thermal dynamics of a negatively-buoyant spherical colloidal particle between two rigid walls, including surface charges. After recalling the overdamped Langevin equation including spurious forces, we solve it using an optimized numerical scheme, and investigate the full displacement statistics. As a central outcome of our work, we show in particular that special care needs to be taken in order to avoid convergence issues when computing high-order cumulants from Brownian realizations.

\section{Model}
\subsection{Bulk Langevin equation}
We consider a colloidal particle of radius $a$, immersed in a fluid of dynamic shear viscosity $\eta_0$. In the bulk, the particle motion is described by the Langevin equation~\cite{langevinTheorieMouvementBrownien1908}: 
\begin{equation}
	m \ddot{\vec{r}}(t) = - \gamma \dot{\vec{r}}(t) + \vec{\mathcal{F}}(\vec{r}(t)) + \sqrt{2 k_\mathrm{B}T \gamma} ~\vec{w}(t) ~,
	\label{LangevinBrut}
\end{equation}
where $\vec{r}(t) = [r_x(t), r_y(t), r_z(t)]$ is the particle center of mass position at time $t$, $m = \frac{4}{3} \pi a^3\rho$ is the particle mass, $\rho$ is the particle density, $\gamma = 6 \pi \eta_0 a$ is the bulk Stokes drag coefficient, $k_\mathrm{B}$ is the Boltzmann constant, $T$ is the temperature, $\vec{\mathcal{F}}=-\vec{\nabla} V$ is the total conservative force deriving from the potential $V[\vec{r}(t)]$, and $\sqrt{2 k_\mathrm{B}T \gamma} ~\vec{w}(t)$ is the stochastic Langevin force accounting for the random impacts of surrounding fluid molecules. In the following, the projected equations along $x$ and $y$ being independent and similar, we only consider the $x$ axis. The two relevant spatial directions are thus indexed by $i = x, z$, corresponding to the  coordinates  $r_x(t) = x_t$ and $r_z(t) = z_t$. We model $\vec{w}(t) = [w_x(t), w_y(t), w_z(t)]$ as a Gaussian white noise of zero mean $\langle w_i(t) \rangle = 0$, and delta-correlated variance $\langle w_i(t) w_j(t') \rangle = \delta_{ij} \delta(t-t')$, where $\langle \cdot \rangle$ indicates the ensemble average, $\delta_{ij} $ the Kronecker symbol and $ \delta$ the Dirac distribution. 
\begin{figure}[t!]
	\includegraphics[scale=0.8]{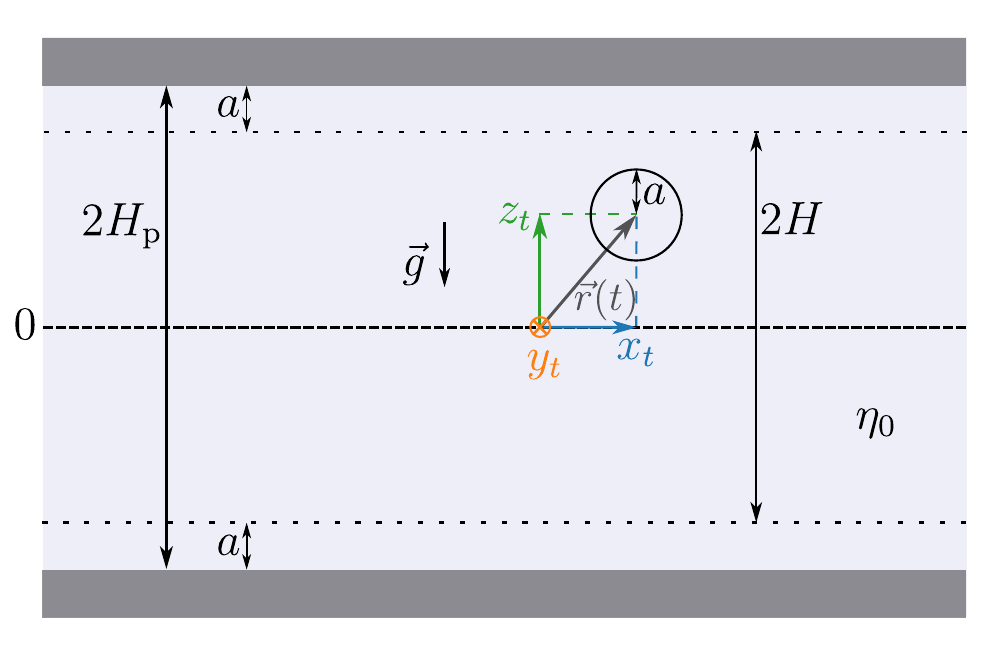}
	\caption{Schematic of the system in the $(x,y)$ plane. A particle of radius $a$ diffuses in three dimensions in a fluid of viscosity $\eta_0$, between two rigid and flat walls separated by a distance $2H_\mathrm{p}=2H+2a$. Besides, and in addition to gravity, we consider specific repulsive screened electrostatic potentials induced by surface charges on the particle and the walls, which are common in experiments~\cite{lavaudStochasticInferenceSurfaceInduced2021} but do not affect the generality of the numerical approach developed here.} 
	\label{schema}
\end{figure}

The inertial term $m \ddot{\vec{r}}$ can be further neglected in the overdamped regime, which is reached when considering times greater than the inertial time scale $m/\gamma \approx 50~\mathrm{ns}$, for $a= 1.5~\mu\mathrm{m}$, $\rho = 1050~\mathrm{kg.m}^{-3}$ and $\eta_0 = 1~\mathrm{mPa.s}$. We note that, for a colloidal particle in a solvent, the underdamped Langevin equation is anyway not necessarily appropriate since there is no clear separation of time scales between hydrodynamic back flow and momentum relaxation~\cite{franoschResonancesArisingHydrodynamic2011}.

\subsection{Overdamped Langevin equation in confinement}
We now consider that the particle is confined between two rigid and flat walls separated by a distance $2 H_\mathrm{p}$, as shown in \cref{schema}. The gravitational acceleration $\vec{g}$ is oriented along $- z$. We further suppose that the particle and walls are negatively charged in water, inducing electrostatic interactions. Due to the presence of a surface-charged particle between two surface-charged planes in a dielectric medium, field reflections on the walls are expected \cite{maxianFastSpectralMethod2021}. These reflections can be seen as electric fields created by the image charges of the particle on the top and bottom walls. Here, we simplify the problem by considering a large-enough gap, so that we can ignore the contribution of such reflections. In addition, the Coulombian electrostatic interactions are screened by the ions present in water. Taking into account gravity and assuming a linear superposition of the Debye-H\"{u}ckel screened electrostatic interactions from each wall, the total potential energy $V(z)$ reads:
\begin{equation}
	\frac{V(z)}{k_\mathrm{B}T}= \displaystyle B \left[ \,\textrm{e}^{-(H+z)/l_\mathrm{D}} + \,\textrm{e}^{-(H-z)/l_\mathrm{D}} \right] + \frac{H+z}{l_\mathrm{B}}\ ,
	\label{Eq:PDF}
\end{equation}
where $B$ is a dimensionless electrostatic magnitude related to the particle and wall surface-charge densities~\cite{behrensChargeGlassSilica2001}, $l_\mathrm{D}$ is the Debye length, $l_\mathrm{B}=k_\mathrm{B}T / (g\Delta m)$ is the Boltzmann length, and $\Delta m = m-\frac{4}{3}\pi a^3 \rho_\mathrm{f} $ is the particle buoyant mass with $\rho_\mathrm{f}$ the fluid density.

Moreover, the presence of the walls modifies the particle mobilities in both the $x$ and $z$ directions, in an anisotropic fashion. Therefore, the Stokes drag coefficients now become space and direction dependent, and we note them $\gamma_{i}(z)=6\pi a\eta_i(z)$, with $\eta_i(z)$ the local effective viscosities. Assuming a linear superposition of the contributions of each wall, one has~\cite{faucheuxConfinedBrownianMotion1994, feitosaWalldragEffectDiffusion1991a}:
\begin{equation}
	\eta_i(z)\simeq \eta^{(1)}_i(H+z)+\eta^{(1)}_i(H-z) - \eta_0~,
	\label{etatot}
\end{equation} 
where we invoked the single-wall expressions $\eta^{(1)}_i$. When neglecting slippage at the wall~\cite{duque-zumajoDiscreteHydrodynamicsSolid2019}, the latter are given by the functional forms~\cite{brennerSlowMotionSphere1961}:
\begin{equation}
	\eta^{(1)}_x(u)= \eta_0\, \frac1{1 - \frac{9}{16}\xi + \frac{1}{8}\xi^3 - \frac{45}{256}\xi^4 - \frac{1}{16}\xi^5}~, 
	\label{eq_para}
\end{equation}
with $\xi = a/(u+a)$, and:
\begin{equation}
	\eta^{(1)}_z(u)=  \eta_0 \,\frac{6u^2 + 9au + 2a^2} {6u^2 + 2au} ~,
\label{eq_perp}
\end{equation}
where the last expression is a Padé approximation \cite{bevanHinderedDiffusionColloidal2000} of the complete formula \cite{brennerSlowMotionSphere1961, faxenBewegungStarrenKugel1923}, valid with less than $1$\% error. We note that, in the horizontal direction $x$, we have omitted the supplementary logarithmic correction for the mobility in the very near vicinity of the wall~\cite{goldmanSlowViscousMotion1967}, as this region is typically not accessed in practice due to the electrostatic repulsion between the particle and the walls~\cite{lavaudStochasticInferenceSurfaceInduced2021}.
 
\begin{figure}[t!]
	\includegraphics[scale=1.]{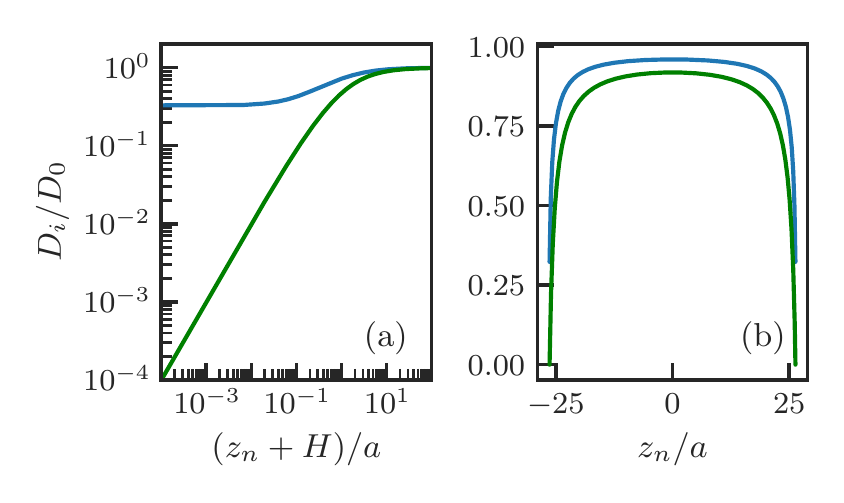}
	\caption{Diffusion coefficients $D_i$ along $x$ (blue) and $z$ (green), normalized by the bulk value $D_0 = k_\mathrm{B}T / (6 \pi a \eta_0 )$\cite{einsteinUberMolekularkinetischenTheorie1905}, as functions of the rescaled  discrete vertical coordinate $z_n/a$, as obtained from Eqs.~(\ref{etatot},\ref{eq_para},\ref{eq_perp},\ref{Dlocal}), with $z_n=z_t$ for $t=n\delta t$. Two typical situations are considered: (a) near the bottom rigid wall, with the particle-wall contact point shifted here to correspond to $z_{n}=0$; (b) between two rigid walls.} 
	\label{Diffusion}
\end{figure}

Invoking the Stokes-Einstein relation, we then construct the local diffusion coefficients, as:
\begin{equation}
	D_i(z) = \frac{ k_\mathrm{B}T }{ \gamma_{i}(z) }~.
	\label{Dlocal}
\end{equation}
For illustration, typical diffusion-coefficient profiles are show in \cref{Diffusion}(a) near the bottom rigid wall, and in \cref{Diffusion}(b) for two rigid walls.

We can then rewrite \cref{LangevinBrut} in the overdamped regime for a particle between two rigid walls, as:
\begin{equation}
	\left\{
	\begin{aligned}
		\frac{\mathrm{d} x_t}{\mathrm{d} t} & = \sqrt{2 D_{x}(z_t)}\, w_x(t) ~, \\
		\frac{\mathrm{d} z_t}{\mathrm{d} t} & = \sqrt{2 D_{z}(z_t)}\, w_z(t) + \frac{\mathcal{F}_z(z_t)}{\gamma_z(z_t)} ~,
	\end{aligned}\right.
\label{langevinModelOverdamped}
\end{equation}
with $ \mathcal{F}_z(z_t)=- V'(z_t)$, where the prime indicates one derivative with respect to the argument. We stress that, since we consider the overdamped regime, we need to specify further the interpretation of the noise. We adopt the Îto convention, and there is thus an additional spurious drift to consider, as explained in details in \cref{section:spuriousdrift}. Eventually, at long time scales, the system must reach equilibrium, and one should recover the canonical Gibbs-Boltzmann distribution in position:
\begin{equation}
	P_\mathrm{eq}(z) = \frac{\textrm{e}^{- \beta V(z)}}{ 	\int_{-H}^{+H} \mathrm{d}z'\, \textrm{e}^{- \beta V(z')}  }~,
	\label{Peq}
\end{equation}
with $\beta = 1 /(k_\mathrm{B}T)$, and using $V(z)$ from \cref{Eq:PDF}.
\subsection{Spurious drift}
\label{section:spuriousdrift}
 One can observe that the noise magnitude $\sqrt{2 D_{i}(z_t)}$ in \cref{langevinModelOverdamped} depends on the random variable $z_t$ itself -- a feature which is thus usually referred to as ``multiplicative noise". It requires a specific treatment in stochastic calculus, the basics of which being recalled hereafter. Let us rewrite the vertical projection of \cref{langevinModelOverdamped} in a differential form, as:
\begin{equation}
	\mathrm{d}z_t = [U(z_t)  + A(z_t) w_z(t)] \mathrm{d}t\ ,
	\label{zlangevinOverdamped}
\end{equation}
with $A(z_t) = \sqrt{2 D_{z}(z_t)}$, and where $U(z_t)$ is an unknown drift velocity at this stage. For integration, we consider a small time interval between $t$ and $t+\tau$. Since there is an intrinsic ambiguity in the evaluation of $A$ on this interval, we introduce a parameter $\alpha  \in [0,1] $ that characterizes the chosen evaluation instant $t + \alpha \tau$. Note that there are two common conventions: i) $\alpha = 0$, \textit{i.e.} the Îto convention; and ii) $\alpha= 1 /2 $, \textit{i.e.} the Stratonovich convention. Imposing the steady state of the associated Fokker-Planck equation to be given by the Gibbs-Boltzmann distribution (see \cref{Peq} for the marginal in position), one eventually gets for all conventions~\cite{mannellaItoStratonovich302012, sanchoBrownianColloidalParticles2011}:
\footnotesize
\begin{equation}
	z_{t+\tau} = z_t +\left[\frac{\mathcal{F}_z(z_t)}{\gamma_z(z_t)}+(1-\alpha) A(z_t) A'(z_t)+A(z_t) w_z(t)\right] \tau ~,
	\label{spurious_eq}
\end{equation}
\normalsize
where we have identified $U(z_t)=\mathcal{F}_z(z_t)/\gamma_z(z_t)+(1-\alpha) A(z_t) A'(z_t)$. By comparison with \cref{langevinModelOverdamped}, we see the appearance of a convention-dependent spurious drift velocity $(1-\alpha) A(z_t) A'(z_t) = (1-\alpha) D_z'(z_t)$. Without this correction brought to the discretized overdamped Langevin equation, the simulated $z_t$ realizations would not satisfy the Gibbs-Boltzmann distribution at long times, which is a necessary condition. In the following, we choose the Îto convention ($\alpha=0$) for the practical numerical integration.

\subsection{Numerical simulations}
We discretize the problem through an Euler scheme, by considering a discrete time $t = n \delta t$, with $n$ a positive integer and $\delta t$ the numerical time step. We write $r_i(t)$ as $r_{i, n}$, $(x_t, z_t)$ as $(x_n, z_n)$, and $w_i(t)$ as $w_{i, n}$. The discrete noises $w_{i, n}$ are chosen as independent Gaussian noises, each with zero mean and $1/\delta t$ variance. Specifically, to generate $w_{i, n}$, we first generate a pair of uniformly-distributed random numbers using the Mersenne-Twister generator~\cite{matsumotoMersenneTwister623dimensionally1998}. Then, we transform the latter pair into a Gaussian-distributed random variable using the Box-Muller algorithm~\cite{scottBoxMullerTransformation2011,leeHardwareGaussianNoise2006}. From the discretization of the horizontal projection of \cref{langevinModelOverdamped}, and from \cref{spurious_eq} for the vertical projection, we get the discrete overdamped Langevin equation in the Îto convention:
\begin{equation}
	\left\{
	\begin{aligned}
		x_{n+1} & = x_n + \sqrt{2 D_x(z_n)} w_{x, n} \delta t  \\
		z_{n+1} & = z_n + [ D'_z(z_n) - \beta D_z(z_n) V'(z_n)  
		\\
		&~~~~~~~~~~~~+ \sqrt{2 D_z(z_n)} w_{z, n} ] \delta t  ~.
	\end{aligned} 
	\right.
	\label{goveq}
\end{equation}

\begin{figure}[t!]
	\includegraphics[scale=1.0]{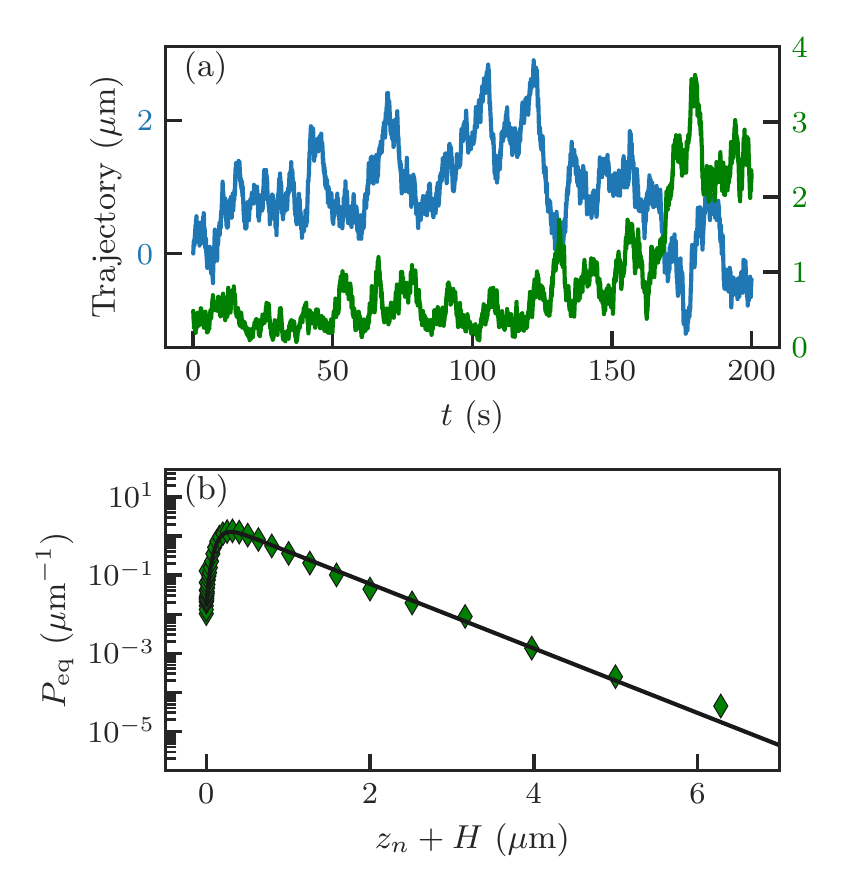}
	\caption{(a) Typical numerically-simulated trajectory of a Brownian particle confined by two rigid walls, in presence of gravity and surface charges. The blue and green lines respectively represent $x_n$ and $z_n + H$, for the first $10^4$ points over a total of $N_\mathrm{t}=10 ^6$ points, using a time step $\delta t = 0.01~\mathrm{s}$. (b) Long-term distribution of the wall-particle distance $z_n+H$. The solid line corresponds to \cref{Eq:PDF,Peq}, with  $a=1.5~\mu \mathrm{m}$, $B=5.0$, $l_\mathrm{B}=526~\mathrm{nm}$, $l_\mathrm{D}=88~\mathrm{nm}$, and $H_{\textrm{p}} = 40 ~\mathrm{\mu m}$.}
	\label{trajectoiry_D_Peq}
\end{figure}

To ensure thermalization in the vertical direction, and avoid unnecessary equilibration delays, we enforce the initial conditions $(x_0, z_0) = (0, z_0)$, with 
$z_0$ randomly sampled from the Gibbs-Boltzmann distribution of \cref{Peq}, using an inverse transformation sampling \cite{devroyeNonUniformRandomVariate1986}.
In the following, each particle trajectory is simulated with $\delta t = 0.01~\mathrm{s}$, $a=1.5~\mu \mathrm{m}$, $\eta_0=1~\mathrm{mPa.s}$, $B=5.0$, $l_\mathrm{B}=526~\mathrm{nm}$, and $l_\mathrm{D}=88~\mathrm{nm}$, in order to reproduce a realistic experimental situation~\cite{lavaudStochasticInferenceSurfaceInduced2021}. A typical trajectory is shown in \cref{trajectoiry_D_Peq}(a), for the $1 000$ first seconds. To make sure that the equilibrium along the vertical direction is reached, we verify that the Gibbs-Boltzmann distribution of \cref{Peq} is reached, without any free parameter, as shown in \cref{trajectoiry_D_Peq}(b). This distribution, together with Eq.~\eqref{Eq:PDF}, provides a direct feeling of the electrostatic depletion near the bottom wall. Without such an electrostatic contribution, the probability of presence would just be a single exponential function, that would be maximal at the wall, as in the historical Perrin's experiments~\cite{perrinMouvementBrownienGrandeurs1909}.

\begin{figure}[t!]
	\includegraphics[scale=0.9]{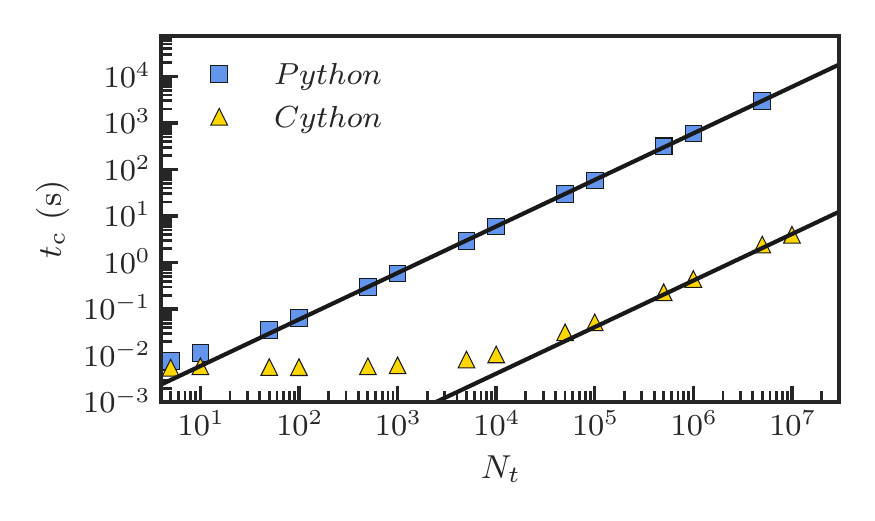}
	\caption{Real computational times $t_\mathrm{c}$ as functions of the total number $N_\mathrm{t}$ of points in a given simulated trajectory, using both \textit{Python} and \textit{Cython}, as indicated. The solid lines correspond to the best linear regressions, from which we find $t_\mathrm{c}(\textrm{s}) = 6 \cdot 10^{-4} N_\mathrm{t}$ for \textit{Python}, and  $t_\mathrm{c}(\textrm{s}) = 4 \cdot 10^{-7} N_\mathrm{t}$ for \textit{Cython}.}
	\label{speedtest}
\end{figure}

We run $N_\mathrm{s}$ simulations. Each simulation produces a trajectory of $N_\mathrm{t}$ points in time. The simulations are performed using \textit{Python} \cite{van1995python}, and each of these can take several seconds of real computation time for $N_\mathrm{t}=10 ^6$, as shown in \cref{speedtest}. For $N_\mathrm{s}=2 \cdot 10^6$ trajectories with $N_\mathrm{t}=10 ^6$, we would need several months of real computation time. To reduce the computational time, we use \textit{Cython} \cite{behnelCythonBestBoth2011}, which allows to keep the flexibility and ease of use of \textit{Python}. As shown in \cref{speedtest}, for $N_\mathrm{t} > 10 ^4$, simulations using \textit{Cython} are a hundred times faster than the ones using \textit{Python}. 

\section{Results}
\label{section:analyse}

\subsection{Mean square displacements}
\begin{figure}[t!]
	\includegraphics{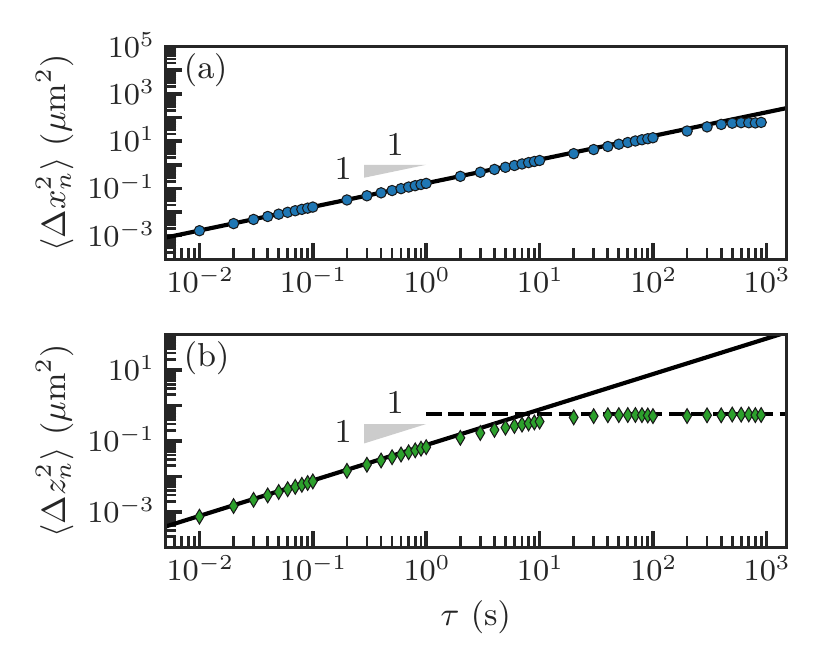}
	\caption{Mean square horizontal (a) and vertical (b) displacements $\langle \Delta r_{i,n}^2 \rangle$ (see \cref{def:MSD}) as functions of time lag $\tau$, for one simulated trajectory of $N_\mathrm{t}=10 ^6$ points, with a numerical time step $\delta t=0.01~\mathrm{s}$. The physical parameters are $a=1.5~\mu \mathrm{m}$, $B=5.0$, $l_\mathrm{B}=526~\mathrm{nm}$, $l_\mathrm{D}=88~\mathrm{nm}$ and $H_{\textrm{p}} = 40 ~\mathrm{\mu m}$. The solid lines correspond to \cref{MSD}, and the dashed line to \cref{pdxlong}.}
	\label{figMSD}
\end{figure}

After having verified above that the simulated system reaches equilibrium properly, one can now turn to the investigation of the dynamical properties of interest. Let us start with the canonical and well-documented quantities, \textit{i.e.} the Mean Squared Displacements (MSDs), which are defined as~\cite{vestergaardEstimationMotilityParameters2015}:
\begin{equation}
	\langle \Delta r_i^2 \rangle (\tau) = \langle [r_i(t+\tau) - r_i(t)]^2 \rangle\ ,
	\label{def:MSD}
\end{equation}
where the ensemble average $ \langle\cdot \rangle$ is computed in practice from  an average $ \langle\cdot \rangle_t$ over time $t$. At all time lags $\tau$ for the horizontal direction, and at small time lags for the vertical one, the MSDs are linear in $\tau$, as shown in \cref{figMSD}. Indeed, the absence of a preliminary ballistic regime is expected for the governing overdamped Langevin equation (see \cref{goveq}). In the bulk, one would have $\langle \Delta r_i^2 \rangle (\tau) = 2D_0\tau$. However, in the confinement situation at stake here, the prefactor is modified. One may expect instead:
\begin{equation}
	\langle \Delta r_i^2 \rangle (\tau)	 = 2 \langle D_i \rangle_0 \tau ~,
	\label{MSD}
\end{equation}
where $\langle \cdot \rangle_0 = \int_{-H}^{+H} \mathrm{d}z\, (\cdot) P_\mathrm{eq}(z)$ is the spatial average over the Gibbs-Boltzmann distribution (see \cref{Peq}). 
Moreover, in \cref{figMSD}(b), one observes that the vertical MSD eventually reaches a plateau at large time lags with a value close to $ l_\mathrm{B}^2$. This saturation corresponds to the fact that the vertical range is limited by gravity, which effectively traps the particle near the bottom wall. The plateau value can be computed from:
\begin{equation}
	\lim_{\tau\rightarrow+\infty}\langle \Delta z_t^2 \rangle =  \int_{-2 H}^{2 H}  \mathrm{d}(\Delta z_t)~\Delta z_t^2 P(\Delta z_t,\tau_{\infty})~ ,
	\label{pdxlong}
\end{equation}
where $P(\Delta z_t, \tau)$ is the Probability Density Function (PDF) of the vertical displacement $\Delta z_t$ at time lag $\tau$, that tends to $P(\Delta z_t, \tau_{\infty})$ when $\tau\rightarrow+\infty$ (see \cref{P_deltaz}) as discussed in the corresponding section. As shown in \cref{figMSD}, \cref{MSD,pdxlong} capture well the numerical data, with no free parameter.

\subsection{Fourth-order cumulants in displacements}
Beyond the MSDs studied in the previous section, \textit{i.e.} the second-order cumulants of the displacements, one can study higher-order cumulants. Such higher-order cumulants -- and especially the horizontal one -- are particularly interesting in order to characterize the inherent non-Gaussianity of the confined Brownian process. The third cumulants $\langle \Delta r_i^3 \rangle_{\textrm{c}}$ are zero, since there is no external drift and $\langle \Delta r_i \rangle = 0$. Therefore, we focus on the fourth cumulants of the displacements:
\begin{equation}
	\begin{split}
		\langle \Delta r_i^4 \rangle_{\textrm{c}} &= \langle \Delta r_i^4 \rangle - 3 \langle \Delta r_i^2 \rangle^2\ .
	\end{split}
	\label{C4}
\end{equation}
For our class of confined systems (see \cref{schema}), in addition to a formal general expression valid at all time lags in the horizontal direction~\cite{alexandreNonGaussianDiffusionSurfaces2022}, one can derive the short-term and long-term asymptotic behaviors of \cref{C4}. At small time lags, one has~\cite{alexandreNonGaussianDiffusionSurfaces2022}:
\begin{equation}
	\langle \Delta r_i^4 \rangle_\mathrm{c} \underrel{\tau \to 0}{\simeq} 12 \left[ \langle D_i^2 \rangle_0 - \langle D_i \rangle_0^2 \right]  \tau ^2\ ,	\label{c4_court}
\end{equation}
where the demonstration for the vertical direction is equivalent to the one for the horizontal direction. We however stress that the non-Gaussianity in the vertical direction is a direct result of vertical confinement, and, as such, is not as profound and interesting as its horizontal counterpart for which the motion is unbounded.  At large time lags, in the horizontal direction, one has~\cite{alexandreNonGaussianDiffusionSurfaces2022}:
\begin{equation}
	\langle \Delta x_t^4 \rangle_\mathrm{c} \underrel{\tau \to +\infty}{\simeq} 24\left(\mathcal{D}_4 \tau - \mathcal{C}_4\right),
	\label{c4_long}
\end{equation}
where $\mathcal{D}_4$ and $\mathcal{C}_4$ are two known constants depending on $V$ and $\{D_i\}$.
At large time lags, in the vertical direction, one expects a plateau given by:
\begin{equation}
	\begin{split}
		\lim_{\tau\rightarrow+\infty}\langle \Delta z_t^4 \rangle &= \int_{-2H}^{+2H}\mathrm{d}(\Delta z_t)\, \Delta z_t^4 P(\Delta z_t, \tau_\infty)  \\
		&~~~ - 3 \left[ \int_{-2H}^{+2H}\mathrm{d}(\Delta z_t)\, \Delta z_t^2 P(\Delta z_t, \tau_\infty) \right]^2  ,
	\end{split}
	\label{c4_eq}
\end{equation}
where $P(\Delta z_t, \tau_\infty)$ is defined in \cref{P_deltaz}, as discussed in the next section.

\begin{figure}[t!]
	\includegraphics{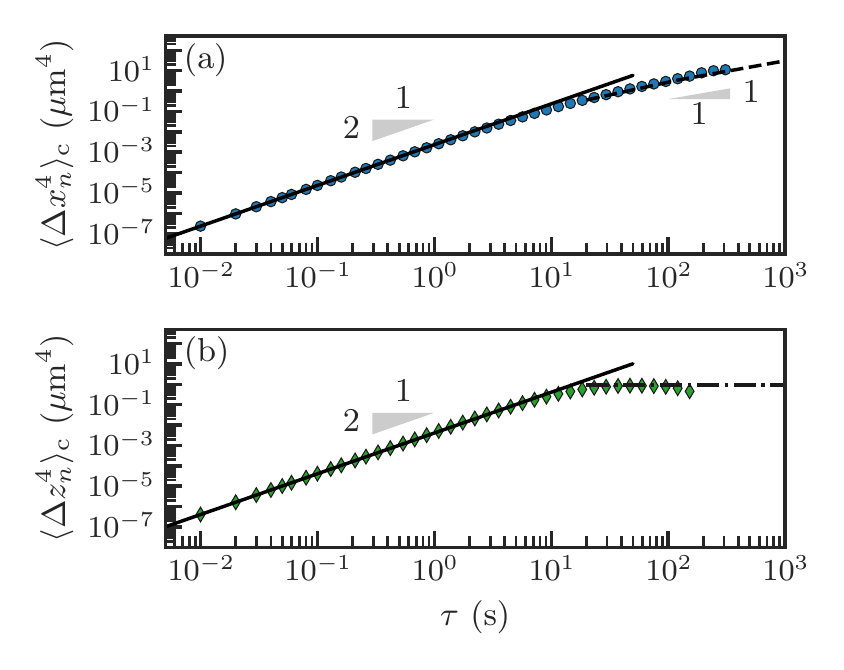}
	\caption{Fourth cumulants $\langle \Delta r_{i,n}^4 \rangle_\mathrm{c}$ (see \cref{C4}) in horizontal (a) and vertical (b) displacements as functions of time lag $\tau$, for $N_\mathrm{s}=2 \cdot 10^6$ simulated trajectory of $N_\mathrm{t}=10 ^6$ points, with a numerical time step $\delta t=0.01~\mathrm{s}$. The physical parameters are $a=1.5~\mu \mathrm{m}$, $B=5.0$, $l_\mathrm{B}=526~\mathrm{nm}$, $l_\mathrm{D}=88~\mathrm{nm}$ and $H_{\textrm{p}} = 40 ~\mathrm{\mu m}$. The solid lines correspond to \cref{c4_court}, the dashed line to \cref{c4_long}, and the dash-dotted line to \cref{c4_eq}.} 
	\label{figC4}
\end{figure}

As shown in \cref{figC4}, the fourth cumulants in displacements obtained from the numerical simulations are in agreement with the asymptotic expressions of \cref{c4_court,c4_long,c4_eq}, with no adjustable parameter. Moreover, we stress that the fourth cumulant in horizontal displacement depends on both $D_x(z)$ and $D_z(z)$ at long times \cite{alexandreNonGaussianDiffusionSurfaces2022}. As such, there is a subtle information coupling between the vertical and horizontal motions, in spite of the fact that the respective noises are not correlated. This is a potentially relevant feature towards the practical extraction of vertical quantities from simple horizontal statistics in actual experimental systems. Note that this idea was already exploited for the second cumulant in a different class of confined systems \cite{Strick1996}.

\subsection{Displacement distributions}
Having discussed the second and fourth cumulants of displacements in the two previous sections, we now turn to the full PDFs $P(\Delta r_i, \tau)$ of displacements $\Delta r_i = r_i(t+\tau) - r_i(t)$, at time lag $\tau$. Note that, in the discretized version for numerical simulations, we denote these quantities $P(\Delta r_{i, n},\tau)$, $\Delta r_{i,n}$, and $\tau=n\delta t$, respectively, with $n$ a positive integer. In the bulk, such PDFs obey the diffusion equation and are classically given by Gaussian distributions, each of zero mean and $2 D_0 \tau$ variance. 

\begin{figure}[t!]
	\includegraphics{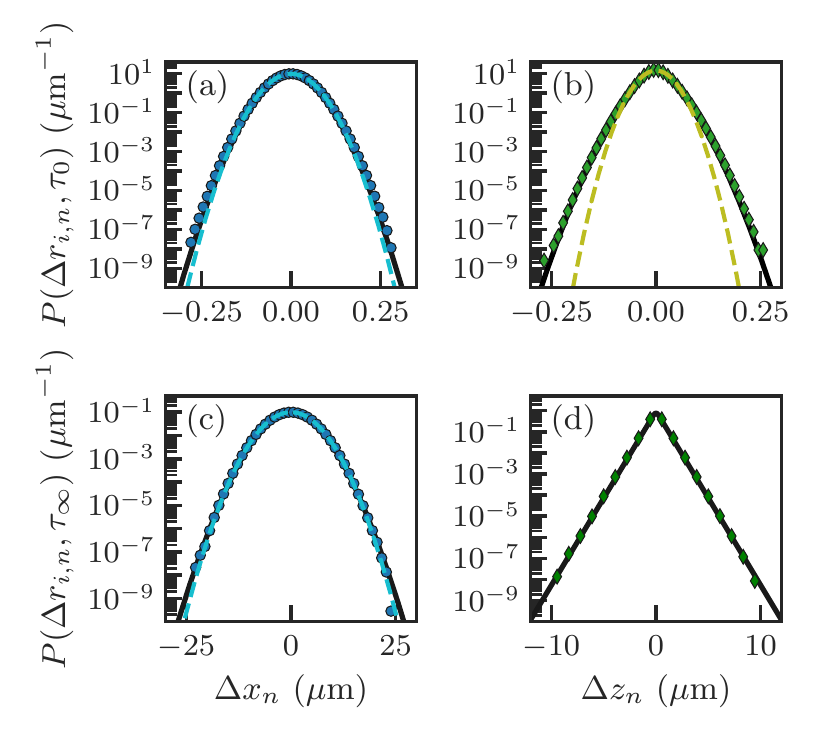}
	\caption{(a,c) Probability density functions $P(\Delta x_n, \tau)$ in horizontal 
displacement $\Delta x_n$, obtained from numerical simulations (blue dots), at small time lag $\tau_0 = 0.01~\mathrm{s}$ and large time time lag $\tau_\infty = 95.4~\mathrm{s}$, respectively. The solid black lines correspond to \cref{P_deltaxi}, and the dashed blue lines correspond to Gaussian distributions of zero means and $2 \langle D_x \rangle \tau$ variances. (b,d) Probability density functions $P(\Delta z_n, \tau)$ in vertical 
displacement $\Delta z_n$, obtained from numerical simulations (green diamonds), at small time lag $\tau_0 = 0.01~\mathrm{s}$ and large time time lag $\tau_\infty = 95.4~\mathrm{s}$, respectively. The solid black lines correspond to \cref{P_deltaxi}, and the dashed green line to a Gaussian distribution of zero mean and $2 \langle D_z \rangle \tau$ variance. In all panels, the PDFs are constructed from $N_\mathrm{s} = 2.1 \times 10^6$ trajectories of $N_\mathrm{t}=10 ^6$ points each, using a numerical time step $\delta t=0.01 ~ \mathrm{s}$, and the physical parameters: $a=1.5~\mu \mathrm{m}$, $B=5.0$, $l_\mathrm{B}=526~\mathrm{nm}$, $l_\mathrm{D}=88~\mathrm{nm}$, and $H_{\textrm{p}} = 40 ~\mathrm{\mu m}$. In all panels, the statistical error bars are smaller than the data symbols.}
	\label{figPDF}
\end{figure}

In our confined case, the presence of the walls modifies the Brownian motion to a so-called Brownian-yet-non-Gaussian motion \cite{han2006brownian,munk2009effective,wang2009anomalous,leptos2009dynamics,wang2012brownian,skaug2013intermittent,chubynsky2014diffusing,guan2014even,jain2016diffusion,jain2016diffusingsurvival,chechkinBrownianNonGaussianDiffusion2017,sposini2018first,lanoiselee2018diffusion,lanoiselee2018model,czajka2019effects,chakraborty2020disorder,hidalgo2020hitchhiker,yin2021non,miotto2021length,alexandreNonGaussianDiffusionSurfaces2022}. We still have a Wiener process, because only the amplitude of the noise is modified by the presence of the walls, but not the Gaussian white noise $w_{i}$(t) itself. The MSDs are linear in time like for the bulk case (see \cref{MSD}). However, the PDFs of displacements are not Gaussian (i.e. with zero mean and $2 \langle D_i(z) \rangle _0$ variance) anymore, in sharp contrast to the bulk case. They depart from Gaussian distributions for large displacements, in particular. At all time lags $\tau$ for the horizontal direction, and at small time lags for the vertical one, the PDFs of displacements can be obtained from spatial averages $\langle\cdot\rangle_0$ of the local diffusion Green's functions over the Gibbs-Boltzmann distribution \cite{lavaudStochasticInferenceSurfaceInduced2021,matseTestDiffusingdiffusivityMechanism2017}: 
\begin{equation}
	P(\Delta r_i, \tau) = \int_{-2H}^{+2H} \mathrm{d}z~ \frac{P_\mathrm{eq}(z)}{\sqrt{4 \pi D_i(z) \tau}} \mathrm{e}^{ - \frac{\Delta r_i^2 }{4 D_i(z) \tau}}  ~.
	\label{P_deltaxi}
\end{equation}
As shown in \cref{figPDF}(a,b,c), the PDFs in displacements obtained from the numerical simulations are in agreement with \cref{P_deltaxi} with no adjustable parameter, at all time lags for the horizontal direction, and at small time lags for the vertical one. Moreover, we observe a departure from the classical bulk Gaussian distributions, that is more pronounced in the vertical direction. Interestingly, even though these three displacement distributions are non-Gaussian, the corresponding MSDs are still linear in time lag (see \cref{figMSD}), as expected for a Brownian-yet-non-Gaussian process.
\begin{figure}[t!]
	\includegraphics[scale=1.0]{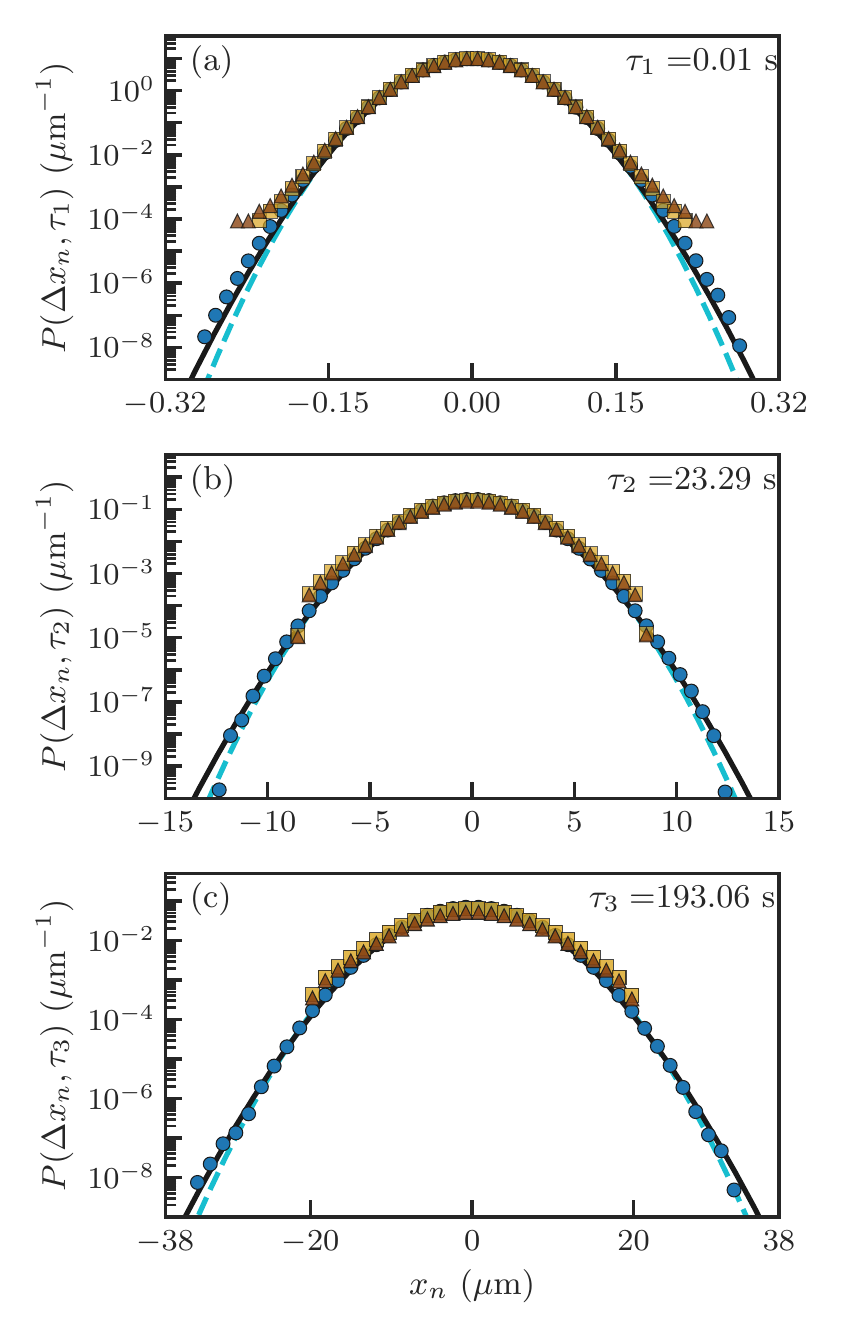}
	\caption{Probability density functions $P(\Delta x_n, \tau)$ in horizontal 
displacement $\Delta x_n$, obtained from numerical simulations (blue dots) by averaging the individual PDFs of the $N_\mathrm{s}=2.1 \times 10^6$ trajectories. Three lag times are considered here: (a) $\tau_1 = 0.01~\mathrm{s}$, (b) $\tau_2 = 1.09~\mathrm{s}$ and (c) $\tau_3 =193.06~\mathrm{s}$. For comparison, are shown the data restricted to within the $90^\mathrm{th}$ (yellow squares) and $99^\mathrm{th}$ (brown triangles) quantiles. The solid black lines correspond to \cref{P_deltaxi}, and the dashed blue lines correspond to Gaussian distributions of zero means and $2 \langle D_x \rangle \tau$ variances. In all panels, the PDFs are constructed from trajectories of $N_\mathrm{t} = 10^6$ points each, using a numerical time step $\delta t=0.01 ~ \mathrm{s}$, and the physical parameters: $a=1.5~\mu \mathrm{m}$, $B=5.0$, $l_\mathrm{B}=526~\mathrm{nm}$, $l_\mathrm{D}=88~\mathrm{nm}$, and $H_{\textrm{p}} = 40 ~\mathrm{\mu m}$.}
	\label{PDF_convergence}
\end{figure}
\begin{figure}[ht!]
	\includegraphics[scale=0.9]{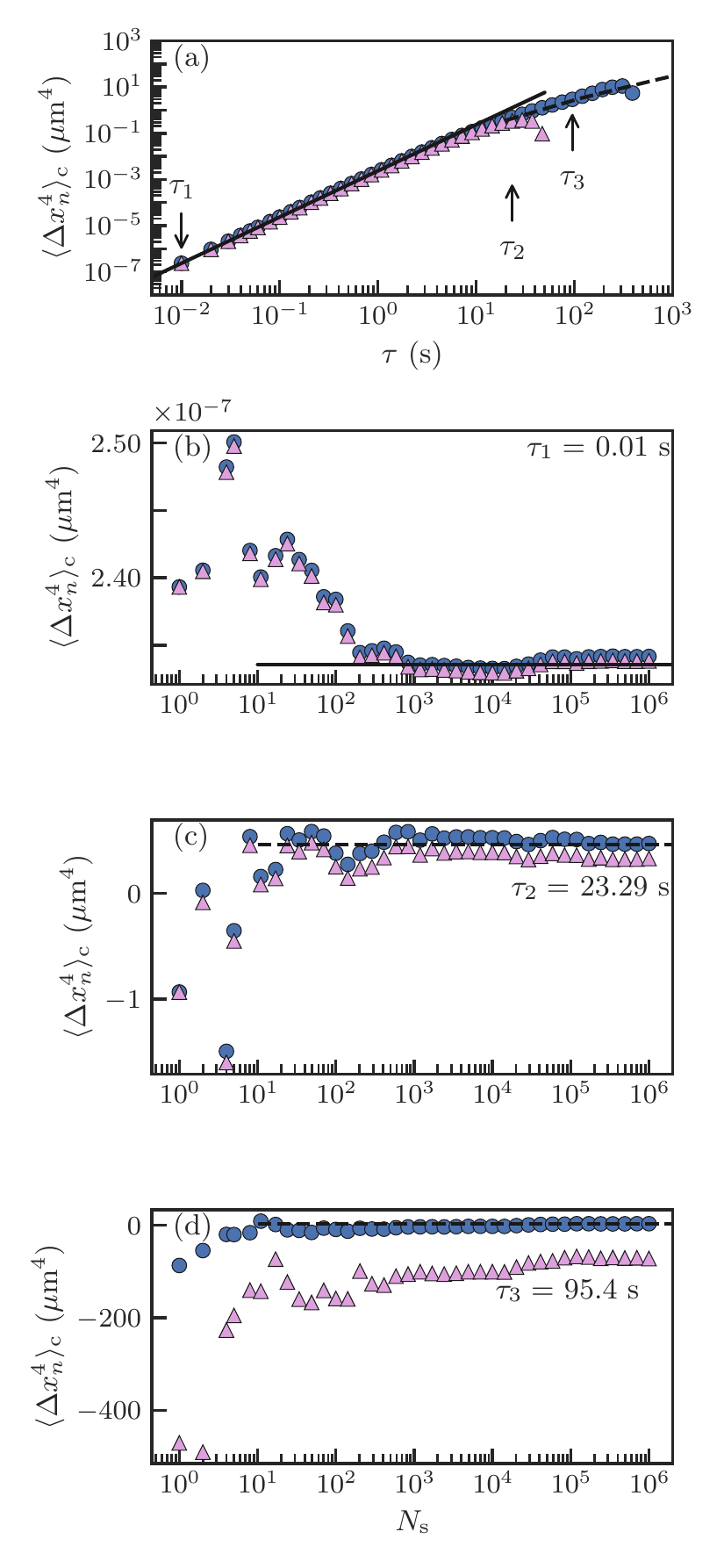}
	\caption{(a) Fourth cumulant $\langle \Delta x_n^4 \rangle_\mathrm{c}$ in horizontal displacement $ \Delta x_n$ as a function of time lag $\tau$, as obtained from the distribution method of \cref{C4_simu,def_calcul_PDx} (blue circles). For comparison, we also show simple averages (pink triangles) of the fourth cumulants in horizontal displacement obtained from the individual trajectories. (b-d) Fourth cumulants $\langle \Delta x_n^4 \rangle_\mathrm{c}$ as functions of the number $N_\mathrm{s}$ of simulations, for three different time lags as indicated. In all panels, the solid lines correspond to the short-term asymptotic expression of \cref{c4_court}, and the dashed lines to the long-term asymptotic expression of \cref{c4_long}. The trajectories have $N_\mathrm{t} = 10^6$ points each, and the numerical time step is $\delta t=0.01 ~ \mathrm{s}$. The physical parameters in the simulation are: $a=1.5~\mu \mathrm{m}$, $B=5.0$, $l_\mathrm{B}=526~\mathrm{nm}$, $l_\mathrm{D}=88~\mathrm{nm}$, and $H_{\textrm{p}} = 40 ~\mathrm{\mu m}$.}
	\label{Convergence}
\end{figure}

Let us now turn to the long-term behaviour of the PDF in vertical displacement. As already observed in \cref{figMSD,figC4}, the second and first cumulants of the vertical displacement reach plateau values at long times. This saturation indicates that equilibrium is reached in the vertical direction. Therefore, one can derive the long-term distribution $P(\Delta z_t, \tau_\infty)\equiv\lim_{\tau\rightarrow+\infty}P(\Delta z_t, \tau)$ from the  Gibbs-Boltzmann distribution (see \cref{Peq}), as \cite{lavaudStochasticInferenceSurfaceInduced2021, matseTestDiffusingdiffusivityMechanism2017}:
\begin{equation}
	P(\Delta z_t, \tau_\infty) = \int_{-2H}^{+2H} \mathrm{d}z~ P_\mathrm{eq}(z) P_\mathrm{eq}(z+\Delta z_t)~.
	\label{P_deltaz}
\end{equation}
Stated simply, at equilibrium, a certain displacement $\Delta z_t$ corresponds to having a certain starting point $z$ and the arrival one $z+\Delta z_t$, both being independently distributed according to the Gibbs-Boltzmann distribution, and with a summation over all possible starting points. As shown in \cref{figPDF}(d), the long-term PDF in vertical displacement obtained from the numerical simulations is in agreement with \cref{P_deltaz}, with no adjustable parameter. Moreover, we observe a marked departure from the classical bulk Gaussian distribution. 

\subsection{Rare events and convergence}
As seen in \cref{figPDF}(a,c), resolving the non-Gaussianities in the distribution of the horizontal displacement implies to measure large displacements, which are rare events, and thus require a lot of numerical data. This is illustrated in \cref{PDF_convergence}, where we see that the non-Gaussian data of interest lies outside the $99^\mathrm{th}$ quantiles. As a direct consequence, a single short trajectory does not allow one to resolve the horizontal non-Gaussianities, and thus the fourth cumulant in horizontal displacement (see Supplementary Material of Ref.~\cite{Rupprecht2018} for a similar convergence problem). One possible strategy to overcome this issue would be to generate a much longer trajectory, \textit{e.g.} of $N_\mathrm{t}=10^8$ points, but this would be at the expense of accumulating important numerical errors on the rare events. In order to circumvent such an error accumulation, we instead simulate $N_\mathrm{s} = 2.1 \times 10^6$ shorter trajectories of $N_\mathrm{t}=10^6$ points each. Note that such an issue is however unimportant for the MSD (see \cref{figMSD}) which is dominated by frequent Gaussian-like events. Therefore, the horizontal MSD can be calculated with a single trajectory of $N_\mathrm{t}=10^6$ points, as in \cref{figMSD}.

Another important practical point to consider is the difficulty in registering $N_\mathrm{s} \times N_\mathrm{t}$ points, in order to produce the fourth cumulants at large time lags in \cref{figC4}. To circumvent this issue, we invoke the equivalent expression of the fourth cumulant:
\begin{equation}
	\begin{split}
		\langle \Delta x_t^4 \rangle_\mathrm{c} &= \int_{-2H}^{+2H} \mathrm{d}(\Delta x_t)\, \Delta x_t^4 P(\Delta x_t,\tau)  \\
		&~~~ - 3 \left[  \int_{-2H}^{+2H}\mathrm{d}(\Delta x_t)\, \Delta x_t^2 P(\Delta x_t,\tau) \right]^2\ .
	\end{split}
	\label{C4_simu}
\end{equation}
From this expression, we see that one just needs to construct $P(\Delta x_n,\tau)$ from all the numerical trajectories, in order to evaluate $\langle \Delta x_t^4 \rangle_\mathrm{c}$. The construction of $P(\Delta x_n,\tau)$ is performed by averaging the PDFs $P^{(k)}(\Delta x_n,\tau)$ of horizontal displacements for the individual trajectories (indexed by the integer $k$), as:
\begin{equation}
 P(\Delta x_n,\tau) = \frac{1}{N_\mathrm{s}} \sum_{k=1}^{N_\mathrm{s}} P^{(k)}(\Delta x_n,\tau) ~.
 \label{def_calcul_PDx}
\end{equation}
In \cref{Convergence}(a), we plot the fourth cumulant in horizontal displacement as a function of time lag, as obtained from this distribution method. For comparison, we also plot the fourth cumulant in horizontal displacement estimated by a naive method, which consists in simply averaging the fourth cumulants in horizontal displacement obtained from the individual trajectories. At small time lags, where single trajectories are sufficient, both methods work properly. At large time lags, the distribution method is still robust, while the naive method underestimates the fourth cumulant. This is intimately rooted in the fact that, as $\tau$ increases, single trajectories do not register enough rare events, which are however essential for measuring non-Gaussianities, as discussed above. As shown in \cref{Convergence}(b-d), increasing $N_\mathrm{s}$ does not solve the problem with the naive method, which fails in converging to the good value at large time lags. In contrast, the distribution method converges properly in the considered $N_\mathrm{s}$ and $\tau$ ranges, and is thus more robust.

\section{Conclusion}
We have numerically investigated the Brownian motion of a negatively-buoyant colloidal particle confined between two flat rigid walls, in presence of surface charges. Specifically, we have solved the discretized overdamped Langevin equation, with an appropriate spurious drift. From the generated trajectories, and with specific care provided regarding the slow convergence of high-order cumulants, we have constructed all the relevant statistical observables. From these, we have in particular checked the convergence to equilibrium, and have quantitatively addressed the non-Gaussianity of the process. As such, our method provides efficient, broadrange and quantitative numerical simulations of Brownian motion in confinement, with potential interest for nanophysics and biophysics.

\section*{Acknowledgments}
The authors thank Arthur Alexandre, Nicolas Fares, Yann Louyer, Thomas Gu\'erin and David Dean, for interesting discussions. They acknowledge financial support from the European Union through the European Research Council under EMetBrown (ERC-CoG-101039103) grant. Views and opinions expressed are however those of the authors only and do not necessarily reflect those of the European Union or the European Research Council. Neither the European Union nor the granting authority can be held responsible for them. The authors also acknowledge financial support from the Agence Nationale de la Recherche under EMetBrown (ANR-21-ERCC-0010-01), Softer (ANR-21-CE06-0029), and Fricolas (ANR-21-CE06-0039) grants. Finally, they thank the Soft Matter Collaborative Research Unit, Frontier Research Center for Advanced Material and Life Science, Faculty of Advanced Life Science at Hokkaido University, Sapporo, Japan. 

\section*{Author contribution statement}
Y.A. and T.S. conceived the study. E.M. and M.L. performed the research. E.M. wrote the first draft of the manuscript. All the authors discussed the data and contributed to the writing of the manuscript.

\section*{Data availability statement}
Data produced for this article are available upon reasonable request to the authors.

\bibliographystyle{unsrt}
\bibliography{Millan2022}

\begin{thebibliography}{10}

\bibitem{brownBriefAccountMicroscopical1828}
Robert Brown.
\newblock A brief account of microscopical observations made in the months of
  {{June}}, {{July}} and {{August}} 1827, on the particles contained in the
  pollen of plants; and on the general existence of active molecules in organic
  and inorganic bodies.
\newblock {\em The Philosophical Magazine}, 4(21):161--173, September 1828.

\bibitem{einsteinUberMolekularkinetischenTheorie1905}
Albert Einstein.
\newblock \"uber die von der molekularkinetischen {{Theorie}} der {{W\"arme}}
  geforderte {{Bewegung}} von in ruhenden {{Fl\"ussigkeiten}} suspendierten
  {{Teilchen}}.
\newblock {\em Annalen der physik}, 4, 1905.

\bibitem{sutherlandDynamicalTheoryDiffusion1905}
William Sutherland.
\newblock A dynamical theory of diffusion for non-electrolytes and the
  molecular mass of albumin.
\newblock {\em The London, Edinburgh, and Dublin Philosophical Magazine and
  Journal of Science}, 9(54):781--785, June 1905.

\bibitem{vonsmoluchowskiZurKinetischenTheorie1906a}
M.~{von Smoluchowski}.
\newblock Zur kinetischen {{Theorie}} der {{Brownschen Molekularbewegung}} und
  der {{Suspensionen}}.
\newblock {\em Annalen der Physik}, 326(14):756--780, 1906.

\bibitem{perrinMouvementBrownienGrandeurs1909}
Jean Perrin.
\newblock {Mouvement brownien et grandeurs mol\'eculaires}.
\newblock {\em Le Radium}, 6(12):353--360, 1909.

\bibitem{langevinTheorieMouvementBrownien1908}
Paul Langevin.
\newblock Sur la th\'eorie du mouvement brownien.
\newblock {\em C. R. Acad. Sci. (Paris)}, 146:530--533, 1908.

\bibitem{Genthon2020}
A.~Genthon.
\newblock The concept of velocity in the history of brownian motion.
\newblock {\em The European Physical Journal H}, 45:49, 2020.

\bibitem{Ardourel2022}
V.~Ardourel.
\newblock Brownian motion from a deterministic system of particles.
\newblock {\em Synthese}, 200:0:1, 2022.

\bibitem{brennerSlowMotionSphere1961}
Howard Brenner.
\newblock The slow motion of a sphere through a viscous fluid towards a plane
  surface.
\newblock {\em Chemical Engineering Science}, 16(3-4):242--251, December 1961.

\bibitem{batchelorBrownianDiffusionParticles1976}
G.~K. Batchelor.
\newblock Brownian diffusion of particles with hydrodynamic interaction.
\newblock {\em Journal of Fluid Mechanics}, 74(1):1--29, March 1976.

\bibitem{Bian2016}
X.~Bian, C.~Kim, and G.~E. Karniadakis.
\newblock 111 years of {B}rownian motion.
\newblock {\em Soft Matter}, 12:6331, 2016.

\bibitem{faucheuxConfinedBrownianMotion1994}
Luc~P. Faucheux and Albert~J. Libchaber.
\newblock Confined {{Brownian}} motion.
\newblock {\em Physical Review E}, 49(6):5158--5163, June 1994.

\bibitem{Dufresne2000}
E.~R. Dufresne, T.~M. Squires, M.~P. Brenner, and D.~G. Grier.
\newblock Hydrodynamic coupling of two brownian spheres to a planar surface.
\newblock {\em Physical Review Letters}, 85:3317, 2000.

\bibitem{Felderhof2005}
B.~U. Felderhof.
\newblock Effect of the wall on the velocity autocorrelation function and
  long-time tail of brownian motion.
\newblock {\em Journal of Physical Chemistry B}, 109:21406, 2005.

\bibitem{Joly2006}
L.~Joly, C.~Ybert, and L.~Bocquet.
\newblock Probing the nanohydrodynamics at liquid-solid interfaces using
  thermal motion.
\newblock {\em Physical Review Letters}, 96:046101, 2006.

\bibitem{Eral2010}
H.~B. Eral, J.~M. Oh, D.~van~den Ende, F.~Mugele, and M.~H.~G. Duits.
\newblock Anisotropic and hindered diffusion of colloidal particles in a closed
  cylinder.
\newblock {\em Langmuir}, 26:16722, 2010.

\bibitem{Benichou2013}
O.~B\'enichou, A.~Bodrova, D.~Chakraborty, P.~Illien, A.~Law,
  C.~Mejia-Monasterio, G.~Oshanin, and R.~Voituriez.
\newblock Geometry-induced superdiffusion in driven crowded systems.
\newblock {\em Physical Review Letters}, 111:260601, 2013.

\bibitem{Mo2015}
J.~Mo, A.~Simha, and M.~G. Raizen.
\newblock Broadband boundary effects on brownian motion.
\newblock {\em Physical Review E}, 92:062106, 2015.

\bibitem{matseTestDiffusingdiffusivityMechanism2017}
Mpumelelo Matse, Mykyta~V. Chubynsky, and John Bechhoefer.
\newblock Test of the diffusing-diffusivity mechanism using near-wall colloidal
  dynamics.
\newblock {\em Physical Review E}, 96(4):042604, October 2017.

\bibitem{Gubbiotti2019}
Alberto Gubbiotti, Mauro Chinappi, and Carlo~Massimo Casciola.
\newblock Confinement effects on the dynamics of a rigid particle in a
  nanochannel.
\newblock {\em Phys. Rev. E}, 100:053307, Nov 2019.

\bibitem{Strick1996}
T.~R. Strick, J.-F. Allemand, D.~Bensimon, A.~Bensimon, and V.~Croquette.
\newblock The elasticity of a single supercoiled dna molecule.
\newblock {\em Science}, 271(5257):1835--1837, 1996.

\bibitem{Bockelmann2002}
U.~Bockelmann, Ph. Thomen, B.~Essevaz-Roulet, V.~Viasnoff, and F.~Heslot.
\newblock Unzipping dna with optical tweezers: High sequence sensitivity and
  force flips.
\newblock {\em Biophysical Journal}, 82(3):1537--1553, 2002.

\bibitem{Sainis2007}
S.~K. Sainis, V.~Germain, and E.~R. Dufresne.
\newblock Statistics of particle trajectories at short time intervals reveal
  fn-scale colloidal forces.
\newblock {\em Physical Review Letters}, 99:018303, 2007.

\bibitem{lavaudStochasticInferenceSurfaceInduced2021}
Maxime Lavaud, Thomas Salez, Yann Louyer, and Yacine Amarouchene.
\newblock Stochastic inference of surface-induced effects using brownian
  motion.
\newblock {\em Phys. Rev. Research}, 3:L032011, Jul 2021.

\bibitem{Rieu2021}
M.~Rieu, T.~Vieille, G.~Radou, R.~Jeanneret, N.~Ruiz-Gutierrez, B.~Ducos, J.-F.
  Allemand, and V.~Croquette.
\newblock Parallel, linear, and subnanometric 3d tracking of microparticles
  with stereo darkfield interferometry.
\newblock {\em Science Advances}, 7(6):eabe3902, 2021.

\bibitem{han2006brownian}
Yilong Han, AM~Alsayed, Maurizio Nobili, Jian Zhang, Tom~C Lubensky, and
  Arjun~G Yodh.
\newblock Brownian motion of an ellipsoid.
\newblock {\em Science}, 314(5799):626--630, 2006.

\bibitem{munk2009effective}
Tobias Munk, Felix H{\"o}fling, Erwin Frey, and Thomas Franosch.
\newblock Effective perrin theory for the anisotropic diffusion of a strongly
  hindered rod.
\newblock {\em EPL (Europhysics Letters)}, 85(3):30003, 2009.

\bibitem{wang2009anomalous}
Bo~Wang, Stephen~M Anthony, Sung~Chul Bae, and Steve Granick.
\newblock Anomalous yet brownian.
\newblock {\em Proc. Nat. Acad. Sc.}, 106(36):15160--15164, 2009.

\bibitem{leptos2009dynamics}
Kyriacos~C Leptos, Jeffrey~S Guasto, Jerry~P Gollub, Adriana~I Pesci, and
  Raymond~E Goldstein.
\newblock Dynamics of enhanced tracer diffusion in suspensions of swimming
  eukaryotic microorganisms.
\newblock {\em Phys. Rev. Lett.}, 103(19):198103, 2009.

\bibitem{wang2012brownian}
Bo~Wang, James Kuo, Sung~Chul Bae, and Steve Granick.
\newblock When brownian diffusion is not gaussian.
\newblock {\em Nature materials}, 11(6):481--485, 2012.

\bibitem{skaug2013intermittent}
Michael~J Skaug, Joshua Mabry, and Daniel~K Schwartz.
\newblock Intermittent molecular hopping at the solid-liquid interface.
\newblock {\em Phys. Rev. Lett.}, 110(25):256101, 2013.

\bibitem{chubynsky2014diffusing}
Mykyta~V Chubynsky and Gary~W Slater.
\newblock Diffusing diffusivity: A model for anomalous, yet brownian,
  diffusion.
\newblock {\em Phys. Rev. Lett.}, 113(9):098302, 2014.

\bibitem{guan2014even}
Juan Guan, Bo~Wang, and Steve Granick.
\newblock Even hard-sphere colloidal suspensions display fickian yet
  non-gaussian diffusion.
\newblock {\em ACS nano}, 8(4):3331--3336, 2014.

\bibitem{jain2016diffusion}
Rohit Jain and Kizhakeyil~L Sebastian.
\newblock Diffusion in a crowded, rearranging environment.
\newblock {\em J. Phys. Chem. B}, 120(16):3988--3992, 2016.

\bibitem{jain2016diffusingsurvival}
Rohit Jain and Kizhakeyil~L Sebastian.
\newblock Diffusing diffusivity: survival in a crowded rearranging and bounded
  domain.
\newblock {\em J. Phys. Chem. B}, 120(34):9215--9222, 2016.

\bibitem{chechkinBrownianNonGaussianDiffusion2017}
Aleksei~V. Chechkin, Flavio Seno, Ralf Metzler, and Igor~M. Sokolov.
\newblock Brownian yet {{Non-Gaussian Diffusion}}: {{From Superstatistics}} to
  {{Subordination}} of {{Diffusing Diffusivities}}.
\newblock {\em Physical Review X}, 7(2):021002, April 2017.

\bibitem{sposini2018first}
Vittoria Sposini, Aleksei Chechkin, and Ralf Metzler.
\newblock First passage statistics for diffusing diffusivity.
\newblock {\em J. Phys. A: Math Theor}, 52(4):04LT01, 2018.

\bibitem{lanoiselee2018diffusion}
Yann Lanoisel{\'e}e, Nicolas Moutal, and Denis~S Grebenkov.
\newblock Diffusion-limited reactions in dynamic heterogeneous media.
\newblock {\em Nat. Comm.}, 9(1):4398, 2018.

\bibitem{lanoiselee2018model}
Yann Lanoisel{\'e}e and Denis~S Grebenkov.
\newblock A model of non-gaussian diffusion in heterogeneous media.
\newblock {\em J. Phys. A: Math. Theor.}, 51(14):145602, 2018.

\bibitem{czajka2019effects}
Pawe{\l} Czajka, Jan~M Antosiewicz, and Maciej D{\l}ugosz.
\newblock Effects of hydrodynamic interactions on the near-surface diffusion of
  spheroidal molecules.
\newblock {\em ACS omega}, 4(16):17016--17030, 2019.

\bibitem{chakraborty2020disorder}
Indrani Chakraborty and Yael Roichman.
\newblock Disorder-induced fickian, yet non-gaussian diffusion in heterogeneous
  media.
\newblock {\em Phys. Rev. Res.}, 2(2):022020, 2020.

\bibitem{hidalgo2020hitchhiker}
M~Hidalgo-Soria and E~Barkai.
\newblock Hitchhiker model for laplace diffusion processes.
\newblock {\em Phys. Rev. E}, 102(1):012109, 2020.

\bibitem{yin2021non}
Qingqing Yin, Yunyun Li, Fabio Marchesoni, Subhadip Nayak, and Pulak~K Ghosh.
\newblock Non-gaussian normal diffusion in low dimensional systems.
\newblock {\em Frontiers of Physics}, 16(3):1--14, 2021.

\bibitem{miotto2021length}
Jos{\'e}~M Miotto, Simone Pigolotti, Aleksei~V Chechkin, and S{\'a}ndalo
  Rold{\'a}n-Vargas.
\newblock Length scales in brownian yet non-gaussian dynamics.
\newblock {\em Phys. Rev. X}, 11(3):031002, 2021.

\bibitem{alexandreNonGaussianDiffusionSurfaces2022}
Arthur Alexandre, Maxime Lavaud, Nicolas Fares, Elodie Millan, Yann Louyer,
  Thomas Salez, Yacine Amarouchene, Thomas Gu{\'e}rin, and David~S. Dean.
\newblock Non-{{Gaussian}} diffusion near surfaces.
\newblock {\em Physical Review Letters}, 2023.

\bibitem{Marbach2018}
S.~Marbach, D.~S. Dean, and L.~Bocquet.
\newblock Transport and dispersion across wiggling nanopores.
\newblock {\em Nature Physics}, 14:1108, 2018.

\bibitem{Sarfati2021}
R.~Sarfati, C.~P. Calderon, and D.~K. Schwartz.
\newblock Enhanced diffusive transport in fluctuating porous media.
\newblock {\em ACS Nano}, 15:7392, 2021.

\bibitem{Benichou2008}
O.~B\'enichou, C.~Loverdo, M.~Moreau, and R.~Voituriez.
\newblock Optimizing intermittent reaction paths.
\newblock {\em Physical Chemistry Chemical Physics}, 10:7059, 2008.

\bibitem{Boniello2018}
Giuseppe Boniello, Christophe Tribet, Emmanuelle Marie, Vincent Croquette, and
  Dra\ifmmode~\check{z}\else\v{z}\fi{}en Zanchi.
\newblock Rolling and aging in temperature-ramp soft adhesion.
\newblock {\em Phys. Rev. E}, 97:012609, Jan 2018.

\bibitem{Alexandre2022}
A.~Alexandre, M.~Mangeat, T.~Gu\'erin, and D.~S. Dean.
\newblock How stickiness can speed up diffusion in confined systems.
\newblock {\em Phys. Rev. Lett.}, 128:210601, May 2022.

\bibitem{Bickel2006}
T.~Bickel.
\newblock Brownian motion near a liquid-like membrane.
\newblock {\em European Physical Journal E}, 20:379, 2006.

\bibitem{Wang2009}
G.~M. Wang, R.~Prabhakar, and E.~M. Sevick.
\newblock Hydrodynamic mobility of an optically trapped colloidal particle near
  fluid-fluid interfaces.
\newblock {\em Physical Review Letters}, 103:248303, 2009.

\bibitem{Daddi2016}
A.~Daddi-Moussa-Ider, A.~Guckenberger and S.~Gekle.
\newblock Particle mobility between two planar elastic membranes: Brownian
  motion and membrane deformation.
\newblock {\em Physics of Fluids}, 28:071903, 2016.

\bibitem{Bar2017}
Chen Bar-Haim and Haim Diamant.
\newblock Correlations in suspensions confined between viscoelastic surfaces:
  Noncontact microrheology.
\newblock {\em Phys. Rev. E}, 96:022607, Aug 2017.

\bibitem{Cejas2017}
C.~M. Cejas, F.~Monti, M.~Truchet, J.-P. Burnouf, and P.~Tabeling.
\newblock Particle deposition kinetics of colloidal suspensions in
  microchannels at high ionic strength.
\newblock {\em Langmuir}, 33:6471, 2017.

\bibitem{Vilquin2021}
Alexandre Vilquin, Vincent Bertin, Pierre Soulard, Gabriel Guyard, Elie
  Rapha\"el, Fr\'ed\'eric Restagno, Thomas Salez, and Joshua~D. McGraw.
\newblock Time dependence of advection-diffusion coupling for nanoparticle
  ensembles.
\newblock {\em Phys. Rev. Fluids}, 6:064201, Jun 2021.

\bibitem{Vilquin2022}
Alexandre Vilquin, Vincent Bertin, Elie Rapha\"el, David~S. Dean, Thomas Salez,
  and Joshua~D. McGraw.
\newblock Nanoparticle taylor dispersion near charged surfaces with an open
  boundary.
\newblock {\em Phys. Rev. Lett.}, 130:038201, Jan 2023.

\bibitem{volpeSimulationBrownianParticle2013}
Giorgio Volpe and Giovanni Volpe.
\newblock Simulation of a {{Brownian}} particle in an optical trap.
\newblock {\em American Journal of Physics}, 81(3):224--230, March 2013.

\bibitem{petersEfficientBrownianDynamics2002b}
E.~A. J.~F. Peters and Th. M. A. O.~M. Barenbrug.
\newblock Efficient brownian dynamics simulation of particles near walls. i.
  reflecting and absorbing walls.
\newblock {\em Phys. Rev. E}, 66:056701, Nov 2002.

\bibitem{simonninDiffusionConfinementHydrodynamic2017}
Pauline Simonnin, Beno\^it Noetinger, Carlos Nieto-Draghi, Virginie Marry, and
  Benjamin Rotenberg.
\newblock Diffusion under confinement: Hydrodynamic finite-size effects in
  simulation.
\newblock {\em Journal of Chemical Theory and Computation}, 13(6):2881--2889,
  June 2017.
\newblock Publisher: American Chemical Society.

\bibitem{sprinkleDrivenDynamicsDense2020}
Brennan Sprinkle, Ernest B. van~der Wee, Yixiang Luo, Michelle M. Driscoll,
  and Aleksandar Donev.
\newblock Driven dynamics in dense suspensions of microrollers.
\newblock {\em Soft Matter}, 16(34):7982--8001, 2020.
\newblock Publisher: Royal Society of Chemistry.

\bibitem{knowlesCurrentFluctuationsNanopores2021}
Stuart~F. Knowles, Nicole~E. Weckman, J.-Y. Lim.~Vincent, Douwe~J. Bonthuis,
  Ulrich~F. Keyser, and Alice~L. Thorneywork.
\newblock Current fluctuations in nanopores reveal the polymer-wall adsorption
  potential.
\newblock {\em Physical Review Letters}, 127(13):137801, September 2021.
\newblock Publisher: American Physical Society.

\bibitem{marbachIntrinsicFractionalNoise2021}
S.~Marbach.
\newblock Intrinsic fractional noise in nanopores: The effect of reservoirs.
\newblock {\em The Journal of Chemical Physics}, 154(17):171101, May 2021.
\newblock Publisher: American Institute of Physics.

\bibitem{franoschResonancesArisingHydrodynamic2011}
Thomas Franosch, Matthias Grimm, Maxim Belushkin, Flavio~M. Mor, Giuseppe
  Foffi, Lászl\'o Forr\'o, and Sylvia Jeney.
\newblock Resonances arising from hydrodynamic memory in {Brownian} motion.
\newblock {\em Nature}, 478(7367):85--88, October 2011.
\newblock Number: 7367 Publisher: Nature Publishing Group.

\bibitem{maxianFastSpectralMethod2021}
Ondrej Maxian, Raúl~P. Peláez, Leslie Greengard, and Aleksandar Donev.
\newblock A fast spectral method for electrostatics in doubly periodic slit
  channels.
\newblock {\em The Journal of Chemical Physics}, 154(20):204107, 2021.

\bibitem{behrensChargeGlassSilica2001}
Sven~H. Behrens and David~G. Grier.
\newblock The charge of glass and silica surfaces.
\newblock {\em The Journal of Chemical Physics}, 115(14):6716--6721, October
  2001.

\bibitem{feitosaWalldragEffectDiffusion1991a}
M.~I.~M. Feitosa and O.~N. Mesquita.
\newblock Wall-drag effect on diffusion of colloidal particles near surfaces:
  {{A}} photon correlation study.
\newblock {\em Physical Review A}, 44(10):6677--6685, November 1991.

\bibitem{duque-zumajoDiscreteHydrodynamicsSolid2019}
D.~Duque-Zumajo, J.~A. de~la Torre, Diego Camargo, and Pep Español.
\newblock Discrete hydrodynamics near solid walls: Non-markovian effects and
  the slip boundary condition.
\newblock {\em Physical Review E}, 100(6):062133, December 2019.
\newblock Publisher: American Physical Society.

\bibitem{bevanHinderedDiffusionColloidal2000}
Michael~A. Bevan and Dennis~C. Prieve.
\newblock Hindered diffusion of colloidal particles very near to a wall:
  {{Revisited}}.
\newblock {\em The Journal of Chemical Physics}, 113(3):1228--1236, July 2000.

\bibitem{faxenBewegungStarrenKugel1923}
Hilding Faxen.
\newblock Die {{Bewegung}} einer starren {{Kugel}} langs der {{Achse}} eines
  mit zaher {{Flussigkeit}} gefullten {{Rohres}}.
\newblock {\em Arkiv for Matemetik Astronomi och Fysik}, 17:1--28, 1923.

\bibitem{goldmanSlowViscousMotion1967}
A.~J. Goldman, R.~G. Cox, and H.~Brenner.
\newblock Slow viscous motion of a sphere parallel to a plane wall—ii couette
  flow.
\newblock {\em Chemical Engineering Science}, 22(4):653--660, April 1967.

\bibitem{mannellaItoStratonovich302012}
Riccardo Mannella and Peter V.~E. McClintock.
\newblock It\^o versus stratonovich: 30 years later.
\newblock {\em Fluctuation and Noise Letters}, 11(01):1240010, March 2012.

\bibitem{sanchoBrownianColloidalParticles2011}
J.~M. Sancho.
\newblock Brownian colloidal particles: {{Ito}}, {{Stratonovich}}, or a
  different stochastic interpretation.
\newblock {\em Physical Review E}, 84(6):062102, December 2011.

\bibitem{matsumotoMersenneTwister623dimensionally1998}
Makoto Matsumoto and Takuji Nishimura.
\newblock Mersenne twister: A 623-dimensionally equidistributed uniform
  pseudo-random number generator.
\newblock {\em ACM Transactions on Modeling and Computer Simulation},
  8(1):3--30, January 1998.

\bibitem{scottBoxMullerTransformation2011}
David~W. Scott.
\newblock Box\textendash{{Muller}} transformation.
\newblock {\em WIREs Computational Statistics}, 3(2):177--179, 2011.

\bibitem{leeHardwareGaussianNoise2006}
D.-U. Lee, J.D. Villasenor, W.~Luk, and P.H.W. Leong.
\newblock A hardware {{Gaussian}} noise generator using the {{Box-Muller}}
  method and its error analysis.
\newblock {\em IEEE Transactions on Computers}, 55(6):659--671, June 2006.

\bibitem{devroyeNonUniformRandomVariate1986}
Luc Devroye.
\newblock {\em Non-{{Uniform Random Variate Generation}}}.
\newblock {Springer}, {New York, NY}, 1986.

\bibitem{van1995python}
Guido Van~Rossum and Fred~L Drake~Jr.
\newblock {\em Python Tutorial}.
\newblock {Centrum voor Wiskunde en Informatica Amsterdam, The Netherlands},
  1995.

\bibitem{behnelCythonBestBoth2011}
Stefan Behnel, Robert Bradshaw, Craig Citro, Lisandro Dalcin, Dag~Sverre
  Seljebotn, and Kurt Smith.
\newblock Cython: {{The Best}} of {{Both Worlds}}.
\newblock {\em Computing in Science \& Engineering}, 13(2):31--39, March 2011.

\bibitem{vestergaardEstimationMotilityParameters2015}
C.~L. Vestergaard, J.~N. Pedersen, K.~I. Mortensen, and H.~Flyvbjerg.
\newblock Estimation of motility parameters from trajectory data.
\newblock {\em The European Physical Journal Special Topics},
  224(7):1151--1168, July 2015.

\bibitem{Rupprecht2018}
J.-F. Rupprecht, A.~Singh~Vishen, G.~V. Shivashankar, M.~Rao, and J.~Prost.
\newblock Maximal fluctuations of confined actomyosin gels: Dynamics of the
  cell nucleus.
\newblock {\em Phys. Rev. Lett.}, 120:098001, Mar 2018.

\end{thebibliography}
\end{document}